\documentclass[twocolumn]{el-author}
\usepackage[noadjust]{cite}

\newcommand{\ua}{\uparrow}
\newcommand{\nc}{\newcommand}
\nc{\da}{\downarrow} \nc{\hc}{\hat{c}} \nc{\hS}{\hat{S}}
\nc{\bra}{\langle} \nc{\ket}{\rangle} \nc{\eq}{equation (\ref}
\nc{\h}{\hat} \nc{\hT}{\h{T}}\nc{\be}{\begin{eqnarray}}
\nc{\ee}{\end{eqnarray}}\nc{\rd}{\textrm{d}}\nc{\e}{eqnarray}\nc{\hR}{\hat{R}}\nc{\Tr}{\mathrm{Tr}}
\nc{\tS}{\tilde{S}}\nc{\tr}{\mathrm{tr}}\nc{\8}{\infty}\nc{\lgs}{\bra\ua,\phi|}\nc{\rgs}{|\ua,\phi\ket}
\nc{\hU}{\hat{U}}\nc{\lfs}{\bra\phi|}\nc{\rfs}{|\phi\ket}\nc{\hZ}{\hat{Z}}\nc{\hd}{\hat{d}}\nc{\mD}{\mathcal{D}}
\nc{\bd}{\bar{d}}\nc{\bc}{\bar{c}}\nc{\mc}{\mathcal}\nc{\ea}{eqnarray}\nc{\mG}{\mathcal{G}}\nc{\bce}{\begin{center}}
\nc{\ece}{\end{center}}
\date{5th February 2016 \small{This paper is a preprint of a paper accepted by Electronics Letters and is subject to Institution of Engineering and Technology Copyright. When the final version is published, the copy of record will be available at IET Digital Library}}

\begin{document}

\title{Binary Polar Codes are Optimized Codes for Bitwise Multistage Decoding}

\author{Mostafa El-Khamy, Hsien-Ping Lin, and Jungwon Lee}

\abstract{Polar codes are considered the latest major breakthrough in coding theory. Polar codes were introduced by Ar{\i}kan in 2008.
In this letter, we show that the binary polar codes are the same as the optimized codes for bitwise multistage decoding (OCBM), which have been discovered before by Stolte in 2002. The equivalence between the techniques used for the constructions and decodings of both codes is established.
}

\maketitle

\section{Introduction}
 Polar codes were introduced by Ar{\i}kan in 2008, who  proved that they can achieve the symmetric capacity of binary-input discrete memoryless channels \cite{arikan2009channel}.
 About seven years earlier, a class of codes called \emph{optimized codes for bitwise multistage decoding} were constructed by Stolte  \cite{stolte2002Phd}. We will refer to the codes of \cite{stolte2002Phd} as OCBM codes. Although they were numerically shown to have near-capacity performance \cite{stolte2002Phd}, they have not received much attention as there was no explicit proof of their capacity achieving property.  Polar codes have received a lot of attention since Ar{\i}kan showed that they have an explicit construction and are provably capacity achieving. The channel polarization phenomenon, introduced by Ar{\i}kan \cite{arikan2009channel}, has since been deployed to show  rate-achievability schemes  in other settings, \emph{cf.} \cite{arikan2010source, abbetelatar12, BICMpolar}.  Many of the ideas used in the construction and decoding of OCBM codes were  based on the works for recursive Plotkin constructions and decodings of Reed-Muller codes \cite{schnabl1995soft, lucas1998improved, stolte2000sequential, stolte2000soft, dumer2000recursive, dumer2001near}. Subsequently, the relationship between Reed-Muller codes and polar codes has also been established \cite{arikan2010survey, PolartoRM}.
 In this paper, we compare between the techniques developed for both polar and OCBM codes, and assert that binary polar codes are the same as the OCBM codes. \cite{GClightning}.

\section{Recursive Plotkin constructions}
The Plotkin construction of two binary outer codes $U$ and $V$ of the same length to give a code $C$ can be described in terms of a generalized concatenation of two codes, i.e. $\mathbf{c}= (\mathbf{u}, \mathbf{u  \oplus v})$, where $\mathbf{c} \in C$,  $\mathbf{u} \in U$, and $\mathbf{v} \in V$. The Plotkin construction can be recursive, where the binary outer codes themselves could have also been obtained by another Plotkin construction. 
Specifically, assuming  that the  initial outer codes are of unit length, then the binary generator matrix of the code constructed by $m$ recursive concatenations is obtained from that of the $(m-1)$-recursively concatenated code by $B_{2^{m}} = \begin{bmatrix}
                               0_{2^{m-1}}  & B_{2^{m-1}} \\
                                B_{2^{m-1}} & B_{2^{m-1}} \\
                              \end{bmatrix} $, where  $B_1 =   \begin{bmatrix}
                                0 & 1 \\
                                1 & 1 \\
                              \end{bmatrix} $.
To construct a code of dimension $k$ and length $N=2^m$, $k$ of the outer codes are chosen to have unit dimension, and the remaining $N-k$ outer codes are chosen to have a zero dimension. Let the information set $\mathcal{I}$ denote the set of indices of the outer codes with unit dimension. If $\mathcal{I}$ is chosen to maximize the minimum Hamming distance of the concatenated code, then this results in a class of codes that includes the binary RM codes \cite{schnabl1995soft, stolte2002Phd}.
\section{OCBM codes}
There have been significant works to improve the performance of RM codes based on its recursive Plotkin construction, \emph{cf.} \cite{schnabl1995soft, lucas1998improved, stolte2000sequential, stolte2000soft, dumer2000recursive, dumer2001near}.  Instead, Stolte \cite{stolte2002Phd} was the first to construct a different code by selecting the rows of its generator matrix from $B_{2^m}$, as indexed by $\mathcal{I}$, in order to minimize the code's word error probability (WEP), rather than to maximize its minimum Hamming distance. The error probabilities of each of the $N$ outer codes are calculated assuming  recursive bitwise  multi-stage decoding (MSD), and $\mathcal{I}$ is chosen to include the indices of the $k$ outer codes with the least error probabilities. Hence, the codes are called optimized codes for bitwise MSD (OCBM). As also mentioned in \cite{stolte2002Phd, arikan2010survey}, related pioneer works were taken by Dumer and Shabunov \cite{dumer2001near} who observed that the decoding performance of RM codes can be improved at the expense of a code-rate loss by using subcodes of RM codes, obtained by setting the coordinates corresponding to the least reliable RM information bits to zero.

For the OCBM construction, one of the methods proposed to estimate the error probabilities of the outer codes uses the sum-capacity observation \cite{stolte2002Phd}. Let $\mathcal{C}_u$ and $\mathcal{C}_v$, respectively, be the channel capacities of the equivalent channels observed by the outer codewords $\mathbf{u}$ and $\mathbf{v}$ with MSD. Let $C_0$ be the symmetric capacity of the binary discrete memoryless channel on which the concatenated codeword $\mathbf{c}$ is transmitted, then $ \mathcal{C}_u + \mathcal{C}_v = 2 \mathcal{C}_0 $.
Assume transmission over a binary input additive white Gaussian noise channel (BIAWGNC)  with signal to noise ratio (SNR)  $\mbox{SNR}_0$ and  channel capacity $\mathcal{C}_0=\mathcal{C}(\mbox{SNR}_0)$,  which can be calculated by numerical integration, \emph{cf.} \cite{john2008digital}. The equivalent SNR (EqSNR) of the equivalent channel observed by $\mathbf{u}$ is $\mbox{SNR}_u = 2  \mbox{SNR}_0$, and of that observed by $\mathbf{v}$ is $\mbox{SNR}_v = \mathcal{C}^{-1}\left( 2   \mathcal{C}(\mbox{SNR}_0) - \mathcal{C}(2  \mbox{SNR}_0)\right)$.  Assuming these two equivalent channels are still Gaussian with the corresponding calculated equivalent SNRs, then the EqSNRs of all $N$ equivalent bit-channels can be calculated by $\log_2(N)$ recursions.
It is clear that $\mathcal{C}_u \geq  \mathcal{C}_0 \geq \mathcal{C}_v$  by the EqSNR method, which is the channel polarization phenomenon later observed by Ar{\i}kan. However, Ar{\i}kan \cite{arikan2009channel} was the first to explicitly prove that the ratio of the number of good bit-channels (whose error probability approaches zero) to the length of the code approaches the channel capacity $\mathcal{C}_0$.  It is worth noting that the sum-capacity observation can also be used to construct OCBM codes for other channels.

 Polar codes \cite{arikan2009channel} were constructed using the Bhattacharryaa parameters (BPs) \cite{kailath1967divergence} as the reliability measure, where the BPs for the binary erasure channel can be recursively calculated to choose the $k$ information bit-channels and freeze the other $N-k$ bit-channels.  Different approaches were later developed to construct polar codes for other channels, notably  the BIAWGNC,  such as the estimation of bit-channel reliability by lower and upper bounds using degrading and upgrading channel quantizations \cite{tal2013construct}, and the density evolution with Gaussian approximation  method (DE-GA) \cite{trifonov2012efficientr}. Similar to the EqSNR method, DE-GA   assumes that the equivalent channels after each polarization step are  Gaussian. Consequently, their corresponding detection log-likelihood ratios (LLRs) are assumed to have symmetric Gaussian distributions whose variance is  twice their mean, i.e.  $\mbox{SNR} = |L_0|/2$, for expected LLR $L_0$.    The DE-GA method  \cite{trifonov2012efficientr} uses the well-known density evolution to track the means of the distributions and
has been generalized to the case of two non-identical input channels with expected LLRs $L_1$ and $L_2$ where the expected LLRs of the degraded and upgraded channels can be respectively calculated by  \cite{elkhamy2015HARQ}:
$|L_v| = \phi^{-1}\left(1-\left(1-\phi(|L_1|)\right)\left(1-\phi(|L_2|)\right)\right)$ and $L_u=L_1+L_2$, such that
\[ \phi(x)= \left\{
 \begin{array}{ll}
 1-\frac{1}{\sqrt{4\pi x}} \int_{-\infty}^\infty \text{tanh}(\frac{u}{2})e^{-\frac{(u-x)^2}{4x}}du, & \mbox{if} \ x >0,\\
 1, & \mbox{if} \ x=0.
 \end{array}\right. \] Similar to the EqSNR method, the equivalent LLRs given by the DE-GA method can be calculated recursively.

\begin{figure}[h]
\vspace{-0.3cm}
\centering{\includegraphics[width=\linewidth, height=3.5cm]{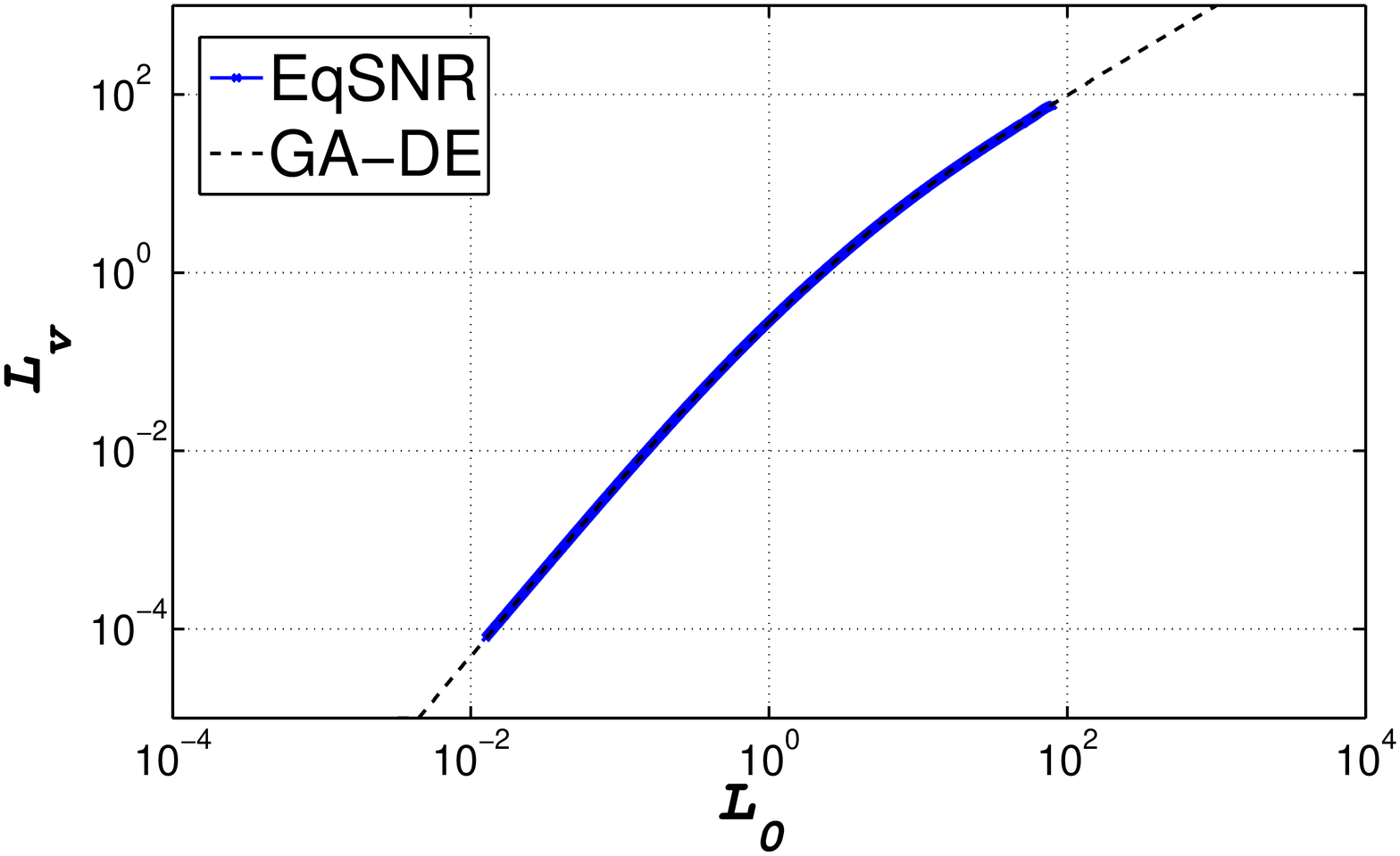}}
\vspace{-0.3cm}
\caption{BIAWGNC Degrading Channel Transfer Function}\label{fig:GAEqSNR}
\vspace{-0.1cm}
\end{figure}

  We implemented both the EqSNR and GA-DE methods using numerical integrations and lookup tables. To construct a code with rate $k/N$,  using either the EqSNR or GA-DE methods,  the $k$ bit-channels with the lowest error probability, or equivalently with the highest EqSNRs or largest expected LLRs,  are chosen to carry the information bits, and the other bit-channels are frozen to zero.
It is clear that in case of identical input LLRs,  the LLRs of the upgraded channels obtained by both methods are twice the input LLRs.
We calculated the relationship between the input LLR ($L_0$) and the estimated LLR of the degraded channel ($L_v$) for the BIAWGNC.  We numerically verify in Fig. \ref{fig:GAEqSNR} that both methods also result in an almost  same degrading channel effect, except that the GA-DE method has better numerical accuracy than the EqSNR method at the extreme ranges due to the capping effect of the capacity functions used to calculate the EqSNR.

Systematic encoding is known to minimize the bit error rate (BER) of binary linear codes under maximum likelihood decoding \cite{Foss98}. Systematic encoding of OCBM codes was done by identifying the set of $k$ independent output code coordinates $\mathcal{O}$  given the set of input indices $\mathcal{I}$, setting $\mathcal{O}$ to carry the desired information bits, erasing the remaining coordinates, and then decoding to recover the erased bits \cite{stolte2002Phd}.  Systematic encoding has also been considered for polar codes by choosing the $k$ output coordinates depending on $\mathcal{I}$,  setting them to the desired information, and recursively solving for the unknown coordinates \cite{arikan2011systematic}. It was also noted in \cite{arikan2011systematic}  that systematic encoding of polar codes can alternatively be done by successive cancellation decoding (SCD) after erasing the unknown coordinates, which is the same as that was adopted for OCBM codes.
Whereas OCBM codes were constructed as generalized concatenated codes, generalized  concatenations of polar codes were  later developed, \emph{cf.}  \cite{trifonov2012efficientr,mahdavifar2014performance, BICMpolar,relaxed}.

\section{Decoding algorithms}
The recursive Plotkin construction of RM codes was used to devise low complexity recursive decoding algorithms. Schnabel and Bossert devised a recursive multistage decoder (MSD) for RM codes based on their recursive Plotkin construction \cite{schnabl1995soft}. Consequently,  an MSD decoder was proposed for OCBM codes \cite{stolte2002Phd}, by deploying the difference of probabilities $h_k$ as a measure of the reliability, where   for an BIAWGNC $h_k = P(c_k=+1|y_k)-P(c_k=-1|y_k) = \tanh\left(L_k/2\right)$, for code bit $c_k$,  channel output $y_k$,  and corresponding channel LLR $L_k$.
 For code length $N$, the reliability of the equivalent bit-channels of the outer code $\mathbf{v}$ are first calculated using the channel observations as $h_k^{(v)} =  h_k \cdot h_{N/2+k}$, from which  the information bits of $\mathbf{v}$ are estimated as $\tilde{v}_k$. The estimated   bits of the outer code $\mathbf{v}$ are then used together with the channel observations to calculate the reliability of bit-channels of the outer code
$\mathbf u$ as $h_k^{(u)} = \frac{h_k + \tilde{v}_k \cdot h_{N/2 + k}}{1 + \tilde{v}_k \cdot h_k \cdot  h_{N/2 + k}}$, from which the information bits of the outer code $\mathbf{u}$ are estimated. This decoder will be referred to as MSD($h$) and can be implemented recursively, where the reliability values calculated at a certain stage are then passed to the next stage till the outer codes are of unit length. Hence, its computational complexity is of  $\mathcal{O}(N \log_2(N))$ \cite{stolte2002Phd}. The successive cancellation decoder proposed for decoding polar codes is similarly recursive with complexity $\mathcal{O}(N \log_2(N))$, but uses the likelihood ratios as reliability values, and will be referred to as SCD($L$) \cite{arikan2009channel}.  Based on the previous works on recursive decoding of RM codes \cite{lucas1998improved, stolte2000sequential, stolte2000soft, dumer2000recursive},  Stolte \cite{stolte2002Phd} also considered sequential stack decoding, list decoding, and list decoding with different permutations of the OCBM codes to improve their finite length decoding performance. Similarly, stack decoding and list decoding have been later considered for decoding polar codes \cite{niu2012stack, tal2011list}. It has been observed that list decoding of OCBM codes has better performance than their stack decoding and approaches their maximum likelihood decoding performance \cite{stolte2002Phd}.

\section{Numerical Comparisons}
\begin{figure}[t]
\centering{\includegraphics[width=\linewidth, height=2.5in]{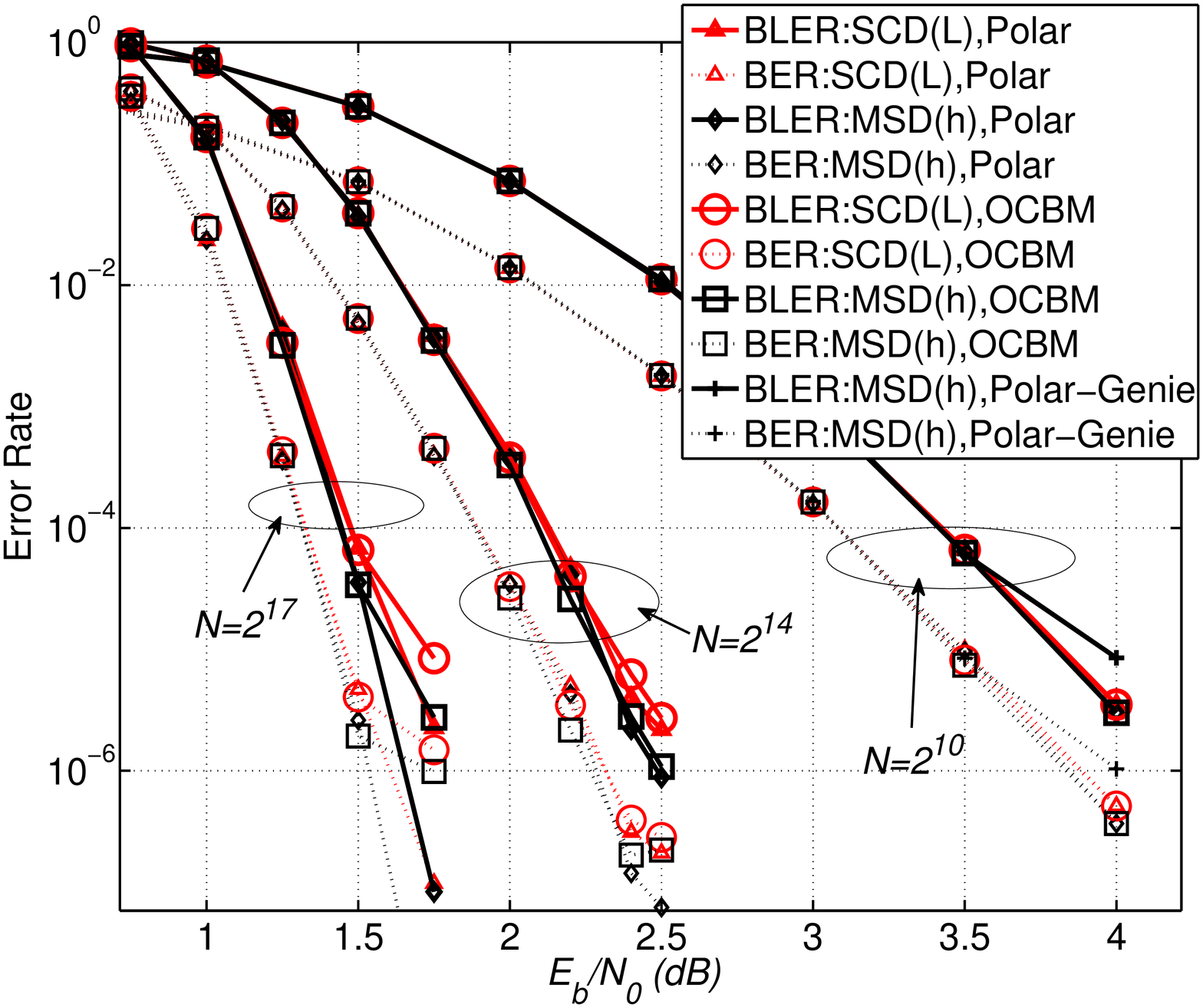}}
\vspace{-0.3cm}
\caption{Polar vs OCBM codes of rate $0.5$, and lengths up to $131,072$ bits }\label{fig:res}
\vspace{-0.3cm}
\end{figure}

In Fig. \ref{fig:res}, we compared the simulated block (BLER) and bit error rate (BER) performances of polar codes and OCBM codes with rate $0.5$ on the BIAWGNC. We show the error-rates at different code lengths $N \in \{ 2^{10}, 2^{14}, 2^{17}\}$.  The codes are reconstructed at each SNR point to illustrate the effectiveness of the construction method. We refer to the codes constructed by the GA-DE method as polar codes, and those constructed by the EqSNR method as OCBM codes. For the code with length $N=2^{10}$, we also show the code constructed by estimating the bit-channel reliability with Monte-Carlo simulations of genie-aided SCD (Genie).  It is observed that all three codes have almost the same performance, except at high SNRs due to numerical inaccuracies. We also compare the performances of the SCD($L$) and MSD($h$) decoders, developed for polar and OCBM codes, respectively. It is observed that their performances are very close, except at higher SNRs and longer codes, where the MSD($h$) decoder gives a better performance by using the difference of probabilities metric.

We conclude that the binary polar codes are the same as the OCBM codes, and the methods developed for decoding both codes are equivalent.

\vskip5pt

\noindent
M. El-Khamy (\textit{Samsung Modem R\&D, San Diego, USA
, and Faculty of Engineering, Alexandria University, Alexandria 21544, Egypt}) \\
H.-P. Lin (\textit{University of California, Davis, CA, USA})\\
Jungwon Lee (\textit{Samsung Modem R\&D, San Diego, USA})
\vskip3pt

\noindent E-mail: m\_elkhamy@ieee.org

\bibliographystyle{IEEEtran}
\bibliography{IEEEabrv,IEEPol2}

\end{document}